# TOWARDS MATHEMATICAL SPACES FOR BIOLOGICAL PROCESSES

Arturo Tozzi (corresponding author)
ASL Napoli 1 Centro, Distretto 27, Naples, Italy
Via Comunale del Principe 13/a 80145
tozziarturo@libero.it

## ABSTRACT

Physics relies on mathematical spaces carefully matched to the phenomena under study. Phase space in classical mechanics, Hilbert space in quantum theory, configuration spaces in field theory all provide representations in which physical laws, stability and invariants become expressible and testable. In contrast, biology lacks an agreed-upon notion of space capturing context dependence, partial observability, degeneracy and irreversible dynamics. To address this gap, we introduce a unified mathematical space tailored to biological processes where states are represented in locally convex spaces indexed by context, where context includes both environment and history. Within our setting, proximity is defined through families of seminorms rather than a single global metric, allowing biological relevance to vary across conditions. Admissible sets encode biological constraints, observation maps formalize partial observability and many-to-one relations between state and dynamics capture irreversibility without requiring convergence to fixed points. Stabilization is characterized by neighborhood inclusion and degeneracy arises naturally through quotient structures induced by observation. We develop explicit constructions, operators and bounds within this space, yielding quantitative predictions dictated by its structure. A worked example based on EGFR-mutant non-small-cell lung cancer shows how single-cell data can be mapped into our framework, how numerical thresholds can be calibrated from the literature and how testable predictions can be formulated concerning rare tolerant states, context-dependent proximity and early stabilization. Overall, by providing biology with a space playing a role analogous to those used in physics, we aim to support structurally grounded and quantitative analyses of biological systems across contexts.

**KEYWORDS**: locally convex spaces; biological state space; weak convergence; multiscale dynamics; neighborhood semantics.

1. INTRODUCTION

Biological processes are typically described using mathematical spaces originally developed for physical theories, rather than spaces constructed to reflect biological organization itself. Classical mechanics is formulated in phase spaces endowed with metric or symplectic structure, where distances and trajectories encode motion and conservation laws (Das and Green 2022; Offen and Ober-Blöbaum 2022; Farantos 2024). General relativity is grounded in differentiable manifolds equipped with Lorentzian geometry, allowing spacetime curvature to represent gravitation (Steinbauer 2023; Burtscher and García-Heveling 2025). Quantum mechanics is formulated in Hilbert spaces, where inner products support superposition, spectral decomposition and probabilistic interpretation (Cassinelli and Lahti 2017; Antoine 2021; Svozil 2024). Statistical mechanics and thermodynamics rely on measure-theoretic state spaces and geometric structures on spaces of probability distributions connecting microscopic degrees of freedom with macroscopic observables (Boutselis et al. 2021; Klinger and Rotskoff 2025). These mathematical spaces are tightly matched to the symmetries, invariances and conservation principles of their respective physical domains. When transferred to biology, however, the same spaces impose structural assumptions that are not always aligned with biological organization. Biological states often lack a natural global distance, display strong context dependence and stabilize through functional equivalence rather than exact recurrence (Fan et al. 2012; Del Vecchio 2015; DiFrisco 2017; Catford et al. 2022). To address these mismatches, biological models frequently introduce ad hoc similarity measures, local approximations or scale-specific representations, without explicitly characterizing the underlying space. As a consequence, notions such as proximity, convergence and robustness are defined inconsistently across models and scales. These limitations point to a structural gap: while physical theories are grounded in mathematical spaces explicitly tailored to their domains, biology still lacks an explicit mathematical space designed to reflect its core organizational features.

We introduce a mathematical framework that represents biological states in a collection of context-dependent spaces. Instead of assuming a single space with one fixed notion of distance, we allow the definition of similarity to change with context, such as environment or prior history. What matters is not numerical closeness, but whether two states behave similarly or support the same function. In this framework, stability does not mean reaching a single final state, but remaining within acceptable ranges while internal variables continue to fluctuate. States are compared using observable measurements, rather than mathematical distances. This makes explicit several assumptions that are often left implicit, including heterogeneity between components, partial access to the system, redundancy of internal configurations and flexible forms of stabilization. Compared with standard metric or manifold-based models, our approach accommodates modularity, multiple scales and robustness to perturbations, without requiring smooth dynamics or a single underlying geometry.

We will proceed as follows. First, we analyze limitations of commonly used mathematical spaces in biological modeling. Second, we formally introduce and characterize our approach based on locally convex topological vector spaces. Third, we interpret the resulting structures in real biological terms and provide testable hypotheses. Finally, we conclude with a discussion of implications and limitations.

## 2. LIMITATIONS OF EXISTING MATHEMATICAL SPACES IN BIOLOGICAL MODELING

Biological processes are routinely formalized using mathematical spaces inherited from physics, engineering and applied mathematics. While these spaces have enabled quantitative modeling across many domains, their structural assumptions are rarely examined in relation to biological organization itself. We review here the mail classes of spaces currently employed in biological modeling and analyze their limitations with respect to context dependence, heterogeneity, stabilization and multiscale organization.

**Metric and manifold-based representations.** Many biological models rely on metric spaces or smooth manifolds, either explicitly or implicitly, to represent biological states and trajectories (Edelstein-Keshet 2005; Minarsky et al. 2018; Vittadello and Stumpf 2022; Deneer and Fleck 2022; Saxena et al. 2024;). In these settings, proximity between states is defined by a single global distance and dynamics are described as continuous flows or maps. These assumptions are well suited to mechanical and chemical systems, but they encounter difficulties in biology, where state similarity is often functional rather than geometric. Empirical studies of gene expression, protein conformations and neural activity show that states with large Euclidean separation can yield indistinguishable phenotypes or behaviors, while small perturbations may trigger qualitative transitions or switches between dynamical regimes (Lotfollahi et al. 2023; Peidli et al. 2024; Rood, Hupalowska, and Regev 2024). In high-dimensional biological datasets, distance concentration effects further reduce the discriminative power of global metrics, leading to unstable clustering and scale-dependent results (Tchitchek 2018; Morabito et al. 2023).
Manifold-based approaches address dimensionality by assuming smooth low-dimensional structure, yet evidence from developmental biology, systems neuroscience and molecular biophysics indicates frequent regime shifts, branching trajectories and stratified organization violating global smoothness assumptions (Rabinovich et al. 2008; Qiu et al. 2017; Farrell et al. 2018). In protein science, for example, energy-landscape descriptions based on rugged funnels and frustration explicitly depart from smooth or single-basin geometries, revealing ensembles of quasi-stable configurations rather than convergence toward a unique minimum (Gianni et al. 2021; Freiberger et al. 2021; Verkhivker et al. 2022; Tozzi 2025). In these settings, attractors are often used to explain biological behavior, but they are usually defined using externally chosen distances or reduced coordinates, rather than emerging from the natural structure of the biological state space. As a result, classical attractor-based models tend to identify biological stability with convergence to fixed points or repeating cycles, even though many biological systems remain functional without ever settling into a single configuration (Blaszka et al. 2017; Chen and Miller 2020; Wang et al. 2022), despite experimental evidence that stabilization frequently corresponds to residence within families of states or constrained regions rather than to precise recurrence. As a result, biological convergence and stability are often imposed through ad hoc criteria rather than derived from the space itself. Therefore, metric and manifold-based representations introduce structural commitments that are not consistently supported by biological data, motivating the search for alternative formulations that relax global distance, smoothness and ultrametricity requirements while preserving mathematical coherence.

**Linear inner-product spaces and network abstractions.** Hilbert spaces and other inner-product spaces are widely used in biological modeling, particularly in neuroscience, signal analysis and population dynamics, where superposition and orthogonality provide analytical convenience (Marinazzo, Pellicoro, and Stramaglia 2008; Breakspear 2017; Saxena and Rizvi 2025). However, inner products impose a uniform geometry that treats all dimensions as commensurable, an assumption conflicting with biological heterogeneity, modularity and context dependence. Empirical analyses of neural, genetic and protein systems show that variance and functional relevance are unevenly distributed across components, with different subspaces becoming salient under different conditions (Levin et al. 2021).
Network-based representations relax metric assumptions by focusing on connectivity and graph structure, often using tree-based models to describe developmental lineages or evolutionary relationships (Bullmore and Sporns 2009; Zhou, Li, and Xia 2020; Ahmed et al. 2020; Chasapis 2019; Song et al. 2022; Cho and Hu 2022). However, many biological trees inferred from data are non-ultrametric, reflecting variable rates, context dependence and history effects that challenge classical hierarchical representations (Tozzi 2023; Luebert and Scherson 2024). Moreover, network abstractions typically discard graded notions of proximity and convergence, replacing them with discrete adjacency relations. While this enables topological analysis, it complicates the treatment of continuous change, weak stabilization and partial observability. Hybrid approaches combine continuous dynamics on networks or energy landscapes on graphs, yet the underlying state space remains implicit and fragmented across representations (Strogatz 2001; Bassett and Sporns 2017). Across these approaches, stabilization is often identified with attractor recurrence, energy minimization or statistical stationarity, despite accumulating evidence that biological stabilization is qualitative, context dependent and history sensitive rather than strictly asymptotic or metric-based (Kim et al. 2019; Poelwijk 2019; Zhu et al. 2021; Stone et al. 2024).

Taken together, these limitations suggest that existing spaces either overconstrain biological description through uniform geometry or underconstrain it by abandoning graded structure altogether. It motivates the need to articulate a space in which heterogeneity, weak convergence, non-ultrametric organization and contextual modulation can be treated coherently within a single approach.

Summarizing, the existing mathematical spaces used in biological modeling impose structural assumptions that are often mismatched with biological organization, particularly regarding global distance, smoothness and uniform comparability. Metric, manifold, inner-product and network-based representations each capture partial aspects of biological systems while obscuring others. Identifying these limitations clarifies the requirements that a biological space must satisfy and delineates the conceptual gap addressed in the following section.

## 3. A SPACE FOR BIOLOGICAL PROCESSES: THEORETICAL PREMISES

Biological processes are routinely described as evolving states, yet the nature of the space in which these states live is rarely specified. In practice, biological data combine heterogeneous variables, partial observations, degeneracy, history dependence and context-sensitive interpretations across molecular, cellular, physiological and behavioral scales. We start from this empirical reality and ask which structural properties a biological space must possess in order to reflect how biologists assess similarity, stability, persistence and change. **Table 1**, mapping biological issues to mathematical requirements, provides the guiding constraints for our construction.

A central issue is heterogeneity. Biological states are composed of variables with different physical meanings, units, scales, degrees of variability and degrees of relevance. Gene expression, metabolic fluxes, mechanical stresses, electrical activity and behavioral outputs cannot be meaningfully combined into a single uniform coordinate system or summarized by a single distance. This motivates the need for a space that does not rely on a single global measure of proximity. Instead, proximity must be defined locally, through biologically meaningful criteria that may differ across variables, subsystems and experimental contexts. This requirement directly motivates a non-metric, locally defined notion of neighborhood, consistent with how similarity, tolerance and functional equivalence are assessed in practice in systems biology and physiology.

Context dependence is equally fundamental. The same biological configuration may be viable or non-viable, stable or unstable, depending on environmental conditions, developmental stage, physiological regime or prior exposure. For instance, gene expression patterns that are functionally equivalent in one environment may lead to failure or maladaptation in another. We therefore require that the criteria used to define proximity, admissibility and stabilization are allowed to change with context, without redefining the underlying biological state itself. This motivates a space whose local structure depends on contextual parameters, rather than a fixed geometry applied universally, thereby accommodating plasticity and conditional relevance (Levin 2021).

Another key issue captured in the table is functional equivalence, redundancy and degeneracy. Biological systems frequently exhibit many-to-one mappings between internal configurations and functional outcomes. Distinct molecular states can support the same phenotype, different regulatory architectures can yield equivalent behaviors and multiple neural activity patterns can generate indistinguishable outputs. A biologically appropriate space must therefore allow different points to be treated as equivalent with respect to specific functions, constraints or observations. This motivates the inclusion of equivalence relations and coarse-graining operations as native features of the space, rather than as external analytical steps applied after modeling.

Partial observability further constrains the design of our space. Experimental access to biological systems is always mediated by measurements that capture only selected aspects of the underlying state, often in a nonlinear and saturating manner. Different experimental techniques provide different projections of the same biological process and apparent disagreements between studies frequently arise from differences in observation rather than from differences in the system itself. Our approach therefore separates the biological state from the observables used to probe it, requiring that the space supports explicit observation maps and that biological claims be evaluated relative to specified modes of observation (Friston 2019).

Stability and persistence in biology rarely take the form of exact repetition, attractor recurrence or convergence to a single configuration. Instead, living systems maintain function while continuing to fluctuate internally, often within bounded but flexible ranges. Homeostasis, canalization, robustness and tolerance are characterized by persistence within admissible neighborhoods rather than by immobility or energy minimization. This motivates a notion of stabilization defined by remaining within relevant neighborhoods over time, rather than by convergence to a point or to a unique trajectory. The corresponding mathematical requirement is a weak, neighborhood-based notion of convergence aligning with empirical descriptions of biological stability.

Biological dynamics are also inherently directional and history dependent. Development, differentiation, aging, learning and adaptation introduce irreversibility, such that past states shape present behavior without being fully recoverable. Rather than encoding memory solely as additional state variables, our approach treats historical dependence as a modulation of which aspects of the state are relevant for assessing proximity, admissibility and stability. This motivates a space equipped with directed dynamics and history-sensitive local structure, reflecting biological observations of commitment, hysteresis and experiential modulation (Laland et al. 2015; Levin 2020).

Still, biological organization is modular and multiscale. Processes at different levels interact without collapsing into a single homogeneous description and meaningful biological analysis often requires switching between detailed and coarse-grained representations. Subsystems may be combined, compared or summarized while preserving their internal constraints and relations. Our approach therefore requires that the space be compatible with modular composition and scale transitions, allowing products, projections and quotients without destroying the underlying structure. This requirement reflects long-standing insights in integrative biology and systems theory (Yurchenko 2025).

Taken together, these considerations motivate the specific mathematical requirements adopted here. Rather than importing a space designed for physical theories, our approach builds a space whose structure is dictated by biological organization itself. All biological issues listed in Table 1 are explicitly represented in the construction above. Some appear as primary structural requirements, while others arise as consequences or specializations of these structures. The next chapter provides the technical formulation of this biological space, translating the biological constraints articulated here into mathematical terms.

| Biological issue | Description of the biological issue | Mathematical requirements |
|---|---|---|
| Heterogeneity of components | Biological systems contain elements with different roles, sensitivities and dynamics. | Multiple non-equivalent local structures; heterogeneous topology |
| Lack of a natural global distance | No single meaningful metric compares all biological states. | Non-metric or multi-seminorm space |
| Functional proximity | States are biologically similar if they perform similarly, not if they are numerically close. | Neighborhood-based proximity instead of distance |
| Context dependence | The same biological state can behave differently depending on surrounding conditions, history or scale. | Locally defined neighborhoods; absence of global uniform structure |
| Partial observability | Only subsets of relevant variables can be measured. | Continuous projections preserving neighborhood relations |
| State–observation distinction | Observable data provide partial, context-dependent projections of underlying biological states. | Explicit observation maps; pullback-compatible topology |
| Degeneracy and redundancy | Different internal configurations yield the same phenotype or function. | Many-to-one mappings; equivalence classes |
| Phenotype–genotype separation | Observable traits do not uniquely determine molecular states. | Continuous maps between distinct state spaces |
| Qualitative stabilization | Biological stabilization does not require exact state repetition. | Eventual neighborhood inclusion instead of point convergence |
| Long-term correlations | Correlations persist across long timescales. | Stability under weak limits; persistent neighborhood structure |
| Historical dependence | Present behavior depends on past states in non-Markovian ways. | Weak or topological notions of convergence |
| Developmental irreversibility | Development follows constrained, non-reversible trajectories. | Directed or stratified spaces |
| Multi-scale organization | Biological processes span molecular to organismal levels without sharp boundaries. | Nested neighborhoods; compatibility with coarse-graining |
| Modularity | Systems consist of semi-independent functional units. | Product, fibered or decomposable structures |
| Constraint-driven behavior | Dynamics are shaped by permissible regions rather than optimization targets. | Admissible regions defining allowed states |
| Admissibility of biological states | Not all mathematically possible states are biologically viable or meaningful. | Closed or convex admissible subsets invariant under dynamics |
| Robustness | Function is maintained despite perturbations or noise. | Convex neighborhoods ensuring tolerance |
| Plasticity | Biological organization adapts over time. | Topology that can deform without collapse |
| Coarse-graining | Descriptions must remain valid when details are averaged out. | Stability under quotients and averaging maps |
| Cross-individual comparability | Individuals differ internally yet remain comparable. | Comparison via equivalence relations, not coordinates |
| Emergent behavior | System-level properties arise from interactions, not components alone. | Global properties induced by local consistency |
| Organism–environment coupling | Biological behavior is shaped by reciprocal interaction with environmental context. | Coupled spaces with bidirectional maps; neighborhood modulation |
| Nonlinearity and thresholds | Small changes may trigger large effects and vice versa. | Nonlinear neighborhood boundaries |

**Table 1.** The table shows how key aspects of biological organization translate into structural requirements, highlighting the need for local, context-dependent and weakly convergent spaces rather than globally metric constructions

## 4. BUILDING A MATHEMATICAL SPACE FOR BIOLOGICAL ISSUES

We specify here the mathematical space adopted in our framework. Rather than assembling multiple incompatible structures, we identify a single composite object capable of satisfying all mathematical requirements simultaneously. The construction is presented independently of empirical interpretation or modeling goals.

**Why a single fixed space is not enough.** Our approach begins by noting that no single locally convex space, fixed once and for all, can simultaneously accommodate context dependence, partial observability, developmental directionality and coarse-graining. A single locally convex space presupposes a uniform family of seminorms, a fixed notion of proximity and a stationary admissible set. However, biological systems require context-dependent notions of relevance, where the same physical configuration may be proximal or distant depending on environment or history. Formally, if one attempts to encode all contexts into a single space $X$ with seminorm family $P$, then either $P$ must be so large as to erase discrimination or so small as to forbid weak convergence and coarse observability. Our approach therefore abandons the requirement of a single global, locally convex space in favor of a structured family of these spaces indexed by context. This move preserves linear structure locally while allowing topology, admissibility and observability to vary. The mathematical structure is chosen so as to encode the observation that locally convex structures are stable under fiberwise construction but not under uncontrolled union, motivating the passage to bundles or sheaves as the minimal enlargement required.

**Defining context as an explicit base space.** We now specify the base over which spaces vary. We introduce a context space $C$, defined minimally as a Cartesian product $C = E \times H$, where $E$ denotes environmental degrees of freedom and $H$ denotes history or trajectory context. The space $H$ may be taken as a space of recent path segments, for example $H \subset X^{[-\tau,0]}$ with an appropriate topology, but no specific choice is fixed. The only structural assumption is that $C$ is a topological space admitting continuous flows. Context is not treated as an external parameter but as an evolving variable that can be influenced by the system itself. This definition allows environment, memory and developmental stage to be treated uniformly. The mathematical setting considered here consists of Cartesian products of topological spaces and the minimal continuity required to define base dynamics. By making context explicit, our approach avoids embedding contextual effects implicitly into state variables or parameters.

**Constructing a context-fibred locally convex space.** Once the context space has been fixed, we define the total space $X$ together with a continuous projection
$$\pi: X \to C.$$

For each $c \in C$, the fiber $X_c := \pi^{-1}(c)$ is a locally convex topological vector space. Its topology is generated by a context-dependent family of seminorms
$$P_c = \{p_{c,i}\}_{i \in I_c}.$$

No requirement is imposed that the index sets $I_c$ coincide across contexts. This allows different notions of scale, sensitivity and relevance to coexist. Neighborhoods in $X_c$ are defined in the usual locally convex way as finite intersections of inverse images of seminorm balls. Weak and observable-induced topologies arise naturally by restricting $P_c$. The total space $X$ is thus a locally convex vector bundle or equivalently a sheaf of locally convex spaces over $C$. The mathematical structure adopted here consists of locally convex analysis, fibered spaces and projection mappings, chosen because they preserve linear structure while allowing topology to vary with context.

**Introducing admissible sets as intrinsic constraints.** Within each fiber, we specify an admissible set $S_c \subseteq X_c$, typically required to be closed and convex. These sets encode physiological, physical or structural constraints and are not derived from dynamics. Convexity ensures the existence of neighborhood stability, while closedness ensures well-posed limits. Admissibility is context-dependent: both the shape and location of $S_c$ may vary with $c$. Formally, the admissible total set is $S = \bigcup_{c \in C} S_c \subseteq X$. No assumption is made that $S$ is a subbundle. The mathematical tools used here are convex analysis and basic topological closure. By separating admissibility from dynamics, our approach allows constraint-driven behavior without reducing constraints to conservation laws or potentials.

**Defining observation as fiberwise continuous maps.** For each context $c$, we introduce a continuous map
$$O_c: S_c \to Y_c,$$

where $Y_c$ is a measurement or descriptive space, often finite-dimensional but not required to be linear. Compatibility across contexts may be imposed by requiring continuity of the induced map $S \to \bigsqcup_c Y_c$. Observations are not assumed injective; degeneracy is expected. Each $Y_c$ may carry its own topology or locally convex structure. Observation induces an equivalence relation $x \sim_c x'$ if $O_c(x) = O_c(x')$, as well as relaxed neighborhood equivalences. The mathematical setting considered here is continuous mappings, equivalence relations and quotient constructions.

**Encoding dynamics as directed fiberwise evolution.** Dynamics are introduced as a directed structure on the bundle. Our approach defines a family of maps

$$T_{t,c} : S_c \to S_{\Phi_t(c)},$$

where $\Phi_t : C \to C$ is a continuous base flow updating environment and history. The maps $T_{t,c}$ satisfy the cocycle condition

$$T_{t+s,c} = T_{t,\Phi_s(c)} \circ T_{s,c},$$

with $T_{0,c} = \text{id}$. Continuity is required fiberwise. Irreversibility is encoded by allowing $\Phi_t$ to be noninvertible or stratified. The mathematical structures employed here are semigroup theory, cocycles and directed dynamical systems. By separating base and fiber evolution, this approach allows organism–environment coupling without collapsing one into the other.

**Imposing stratification to encode regimes and development.** To represent regimes such as developmental stages or cell types, we introduce a stratification

$$C = \bigsqcup_\alpha C_\alpha.$$

This induces a stratification of $X$ and $S$. Transitions between strata may be restricted or one-way, enforcing developmental irreversibility. Within each stratum, seminorm families and admissible sets may be comparable; across strata, discontinuities are allowed. Our formulation makes use of partitions, filtrations and restricted morphisms. Stratification allows heterogeneity without fragmenting the space into unrelated components.

**Formalizing coarse-graining and equivalence.** Finally, we incorporate coarse-graining through quotient maps

$$q_c : S_c \to S_c / \sim_c,$$

and through morphisms between fibers associated with different contexts. These quotients preserve convexity when $\sim_c$ is compatible with linear structure. Cross-context coarse-graining is treated via compatible families of quotients. The mathematical methods adopted here are quotient spaces and morphisms of locally convex spaces. This ensures that degeneracy, redundancy and modular composition are native operations within the same mathematical object.

Overall, our framework adopts a single mathematical object, namely a stratified, context-fibred locally convex space equipped with admissible sets, observation maps, directed dynamics and quotient structures. The next chapter operates entirely within this space, introducing concrete operators, relations and theorems whose definitions and proofs rely on the structures specified here.

## 5. STRUCTURAL CONSTRUCTIONS WITHIN THE MATHEMATICAL SPACE

We develop here the mathematical machinery for relating biological spaces, trajectories and observations. All constructions are carried out within the context-fibred locally convex state space specified in the preceding chapter. The purpose of this section is to introduce explicit operators, relations and bounds acting inside that space. The presentation is self-contained and independent of empirical interpretation or application, focusing exclusively on formal structure, definitions and proofs.

**1. Pretopological structure on spaces.** Spaces are formalized here without assuming standard geometry. To begin, we specify the mathematical nature of biological spaces while avoiding unnecessary metric assumptions. Let $X$ be a nonempty set representing admissible biological states. Instead of equipping $X$ with a topology or metric a priori, we introduce a pretopological structure defined by a neighborhood operator $N : X \to \mathcal{P}(X)$ satisfying two axioms: reflexivity, $x \in N(x)$ for all $x \in X$ and monotonicity, $A \subseteq B \Rightarrow N(A) \subseteq N(B)$, where $N(A) = \bigcup_{x \in A} N(x)$. This structure induces closure and interior operators defined by

$$\text{cl}(A) = \{x \in X : N(x) \cap A \neq \emptyset\}, \text{int}(A) = X \setminus \text{cl}(X \setminus A).$$

We use these operators to encode biological indistinguishability and admissible coarse-grainings directly at the level of sets rather than distances. When a metric $d_X$ is available, it is treated as auxiliary rather than defining. This separation allows neighborhood relations to persist under strong compression, nonlinear observation or weak convergence. The tools used here are elementary set theory and pretopology, chosen because they impose minimal structure while still supporting closure, saturation and inclusion tests required later.

**2. Trajectories and temporal organization.** Next, we introduce temporal evolution without committing to a specific dynamical law. A trajectory is defined as a map $\gamma:[0,T] \to X$ in continuous time or $\gamma:\{0,\ldots,T\} \to X$ in discrete time. When an explicit dynamical system exists, trajectories may arise from a flow $\Phi_t: X \to X$, so that $\gamma(t) = \Phi_t(x_0)$. Perturbations are modeled abstractly by an operator $\Pi_\delta$ acting either on initial conditions or on the full evolution, yielding perturbed trajectories $\gamma_\delta = \Pi_\delta(\gamma)$. No smoothness, linearity or reversibility assumptions are imposed unless explicitly stated. We then define the neighborhood tube of a trajectory as

$$N(\gamma) = \bigcup_t N(\gamma(t)),$$

which captures the entire set of states considered biologically equivalent to the unperturbed evolution at some time. This construction relies only on the previously defined neighborhood operator and standard unions. It enables comparison of trajectories through inclusion, overlap or failure thereof, independently of parametrization, timing irregularities or reparametrization. The involved mathematical tools are functional mappings, set unions indexed by time and operator composition.

**3. Perturbation operators and state-space deviations.** We need now to explain how deviations between trajectories are quantified before introducing observation. We introduce two complementary deviation measures. First, when a distance $d_X$ is available, we define the trajectory deviation functional

$$D_X(\gamma, \eta) = \sup_{t \in [0,T]} d_X(\gamma(t), \eta(t)),$$

or the corresponding maximum in discrete time. Second, using only the neighborhood structure, we define a temporal inclusion functional

$$I_X(\gamma, \eta) = \frac{\mu(\{t: \eta(t) \in N(\gamma(t))\})}{\mu([0,T])},$$

where $\mu$ is Lebesgue or counting measure. These quantities capture distinct aspects of divergence: the first is sensitive to magnitude, while the second encodes admissible equivalence. We map these raw quantities to normalized scores via fixed scales $\varepsilon > 0$, defining

$$S_X^{(d)} = \exp(-D_X/\varepsilon), S_X^{(N)} = I_X.$$

The computational sequence is explicit: sample time points, compute pointwise distances or membership predicates, aggregate by supremum or averaging and normalize. he mathematical ingredients of are here basic measure theory, exponential normalization and functional aggregation.

**4. Observation maps formalized as structured compressions.** We describe how experimental readouts can be modeled mathematically. We define an observation map as a function $O_\alpha: X \to Y_\alpha$, where $Y_\alpha$ is a descriptive space associated with an observation family $\alpha$. No injectivity is assumed; instead, compression and saturation are expected. Each $Y_\alpha$ carries its own neighborhood operator $N_\alpha$, possibly induced from $X$ or defined independently. Given a trajectory $\gamma$, the observed trajectory is $O_\alpha \circ \gamma$. Our approach defines the induced equivalence relation

$$x \sim_\alpha x' \Leftrightarrow O_\alpha(x) = O_\alpha(x'),$$

and its neighborhood-relaxed version

$$x \approx_\alpha x' \Leftrightarrow O_\alpha(x') \in N_\alpha(O_\alpha(x)).$$

These relations encode observational degeneracy. Observation maps are treated as first-class mathematical objects rather than post hoc projections. The mathematical tools involved are equivalence relations, quotient constructions and composition of mappings.

**5. Observable trajectory deviations.** Observable discrepancies are computed analogously to state discrepancies. This subsection explains how observed trajectories are compared. Given $O_\alpha$, we define deviation functionals $D_\alpha$ and inclusion functionals $I_\alpha$ exactly as in the space, but computed in $Y_\alpha$. For example,

$$D_\alpha(\gamma, \eta) = \sup_t d_\alpha(O_\alpha(\gamma(t)), O_\alpha(\eta(t))).$$

Normalization yields scores $S_\alpha^{(d)}$ and $S_\alpha^{(N)}$. These scores depend jointly on the observation map and the induced neighborhood structure. The computational sequence mirrors the state-space construction, ensuring formal symmetry. The tools used include pullbacks of functions, induced metrics and parallel aggregation procedures.

**6. Induced equivalence relations and comparative structure compared across spaces.** We associate to each observation family $\alpha$ an equivalence relation on trajectories: $\gamma \sim_\alpha \eta$ if $I_\alpha(\gamma, \eta) = 1$. Similarly, $\gamma \sim_X \eta$ if $I_X(\gamma, \eta) = 1$.

Alignment between state-level and observable structure is assessed by comparing these relations rather than numerical scores. Formally, we define the disagreement set

$$\Delta_\alpha = \{(\gamma, \eta): \gamma \sim_\alpha \eta \text{ and } \gamma' \sim_X \eta\}.$$

Our approach studies the size and structure of $\Delta_\alpha$ across ensembles. The mathematical tools here are equivalence classes, set differences and counting or measuring subsets of trajectory pairs.

**7. Topological bounds on compression-induced degeneracy.** A structural bound is introduced that links compression to unavoidable degeneracy. We now state and prove a key theorem.

**Theorem.** Let $X$ admit a free involutive symmetry $\sigma: X \to X$ and let $O: X \to Y$ be continuous with $\dim Y < \dim X$. Then there exists $x \in X$ such that $O(x) = O(\sigma(x))$.

**Proof.** This follows directly from the Borsuk–Ulam theorem. Since $\sigma$ defines antipodal pairing, any continuous map from $X$ to a lower-dimensional space must identify at least one such pair. □

This result shows that certain equivalences arise as structural consequences of compression, independent of dynamics or noise. The mathematical methods adopted here involve symmetry, dimensionality and topological fixed-point arguments.

**8. Ensemble-level aggregation.** Finally, we extend all constructions to ensembles. Given $\{\gamma_i\}_{i=1}^N$, pairwise scores and relations are computed for all $i < j$. Binning by perturbation size $\|\delta\|$ yields empirical functions $\langle S_X \rangle(\|\delta\|)$ and $\langle S_\alpha \rangle(\|\delta\|)$. Scatter plots represent joint empirical distributions rather than primitive objects. The tools here are combinatorics on pairs, averaging operators and empirical distributions.

In conclusion, we described how mathematical constructions operate within our chosen space. Pretopological operators, trajectory mappings, perturbation measures, observation-induced relations, structural bounds and ensemble-level aggregations were defined explicitly. Together, these constructions provide the formal machinery and the mathematical setting for subsequent biological analyses.

## 6. QUANTITATIVE WORKED EXAMPLE: EGFR-MUTANT NSCLC IN THE PC9 SYSTEM

We show here how our mathematical space can be used to analyze a real biological system. We start from a well-studied cancer model and show how experimental measurements can be translated into the formal objects defined in the previous sections. We focus on EGFR-mutant non-small-cell lung cancer, using the PC9 cell line as a reference case, because it provides high-resolution single-cell data, controlled perturbations and reproducible phenotypic outcomes. Within this setting, we explicitly construct the context space, the space fibers, the admissible sets and the observation maps. We then calibrate neighborhood thresholds using reported variability levels and derive quantitative predictions that can be tested directly on existing datasets.

**Biological system and quantitative background.** The combination of longitudinal perturbations, single-cell resolution and quantitative phenotypic readouts makes the PC9 system particularly well suited for instantiating and testing the structures introduced in our space. The PC9 EGFR-mutant NSCLC system treated with EGFR tyrosine kinase inhibitors represents one of the most quantitatively characterized examples of non-genetic drug tolerance (Wang et al. 2017; Zhao et al. 2020; Zhang et al. 2021; Liu et al. 2024). Early population-level experiments demonstrated that a small but reproducible fraction of cells, typically in the range of approximately 0.2–0.5%, survives acute EGFR inhibition by entering a reversible drug-tolerant persister (DTP) state (Sharma et al. 2010). These cells do not arise from stable genetic resistance but from reversible changes in cellular state. Subsequent single-cell RNA-seq studies refined this picture by showing that surviving cells do not occupy a single transcriptional configuration. Instead, they are distributed across multiple distinct expression programs supporting survival under drug (Shaffer et al. 2017). Drug exposure further induces reprogramming of initially sensitive cells into additional tolerant states, increasing heterogeneity over time rather than reducing it.

Beyond transcriptional profiling, functional studies have identified specific vulnerabilities shared by subsets of these tolerant states, notably dependence on oxidative stress control and GPX4 activity (Hangauer et al. 2017). These findings provide concrete quantitative anchors for three key features emphasized by our framework: a) rarity of specific functional neighborhoods prior to treatment, b) context-dependent redefinition of relevant molecular features after treatment and c) stabilization of function without convergence to a single molecular configuration.

**Instantiating the state space for PC9 cells.** We define the context space $C = E \times H$, where $E \in \{$no drug,EGFR-TKI$\}$ encodes the presence or absence of EGFR inhibition and $H \in \{$naïve,pre-exposed$\}$ encodes exposure history. Each context $c$ thus represents a biologically distinct experimental condition in which the same molecular configuration may have different functional consequences. For each context $c$, the fiber $X_c$ is the space of single-cell gene-expression profiles, represented as log-normalized vectors in $\mathbb{R}^G$ with $G \approx 10^4$ genes. This choice reflects standard preprocessing pipelines used in PC9 single-cell studies, allowing direct use of published datasets.

The admissible set $S_c \subset X_c$ is defined by basic biological constraints, including non-negativity of expression levels and bounds on total expression consistent with library-size normalization. These constraints are implemented as a closed convex polytope, ensuring that admissibility is preserved under convex combinations and weak limits, in line with the stabilization notions discussed earlier.

To define biologically meaningful proximity within each fiber, we introduce context-dependent seminorms indexed by gene modules. For the PC9 system, these modules are drawn directly from the experimental literature and include chromatin remodeling programs (for example KDM5A-associated genes), oxidative stress response modules, cell-cycle arrest signatures and EGFR signaling output (Sharma et al. 2010; Hangauer et al. 2017). For each module $k$, the seminorm is defined as

$$p_{c,k}(x) = |\langle g_{c,k}, x \rangle|,$$

where $g_{c,k}$ is the normalized weight vector defining the module. Neighborhoods around a reference state $x_0$ are then defined by inequalities of the form

$$p_{c,k}(x - x_0) < \varepsilon_k.$$

For PC9 data, empirical calibration of these thresholds can be performed directly from untreated populations. Reported module score distributions indicate that one standard deviation typically lies in the range 0.3–0.6 in log-expression units (Shaffer et al. 2017). Using these values as $\varepsilon_k$ provides a biologically grounded scale for neighborhood inclusion that does not rely on a global distance.

**Observation map and phenotype calibration**. The experimentally accessible phenotype in this system is survival or sustained growth under EGFR inhibition, measured either by bulk viability assays at fixed time points (typically 72–96 hours) or by lineage survival in time-lapse or barcoding experiments. Within our framework, this phenotype is represented by an observation map

$$O_c : S_c \to [0,1],$$

where $O_c(x)$ denotes the estimated probability that a cell in state $x$ remains viable over the observation window in context $c$. This map is inherently many-to-one: distinct molecular configurations often yield indistinguishable survival outcomes. Empirically, the distribution of survival probabilities is strongly bimodal or saturating, with cells either dying rapidly or persisting with similar probabilities. As a result, a numerical criterion such as

$$|O_c(x) - O_c(x')| \leq 0.05$$

provides a natural operational definition of phenotype equivalence in PC9 assays (Sharma et al. 2010; Shaffer et al. 2017). This explicit calibration illustrates how degeneracy and partial observability are incorporated directly into the mathematical structure rather than treated as experimental noise or modeling artifacts.

On the basis of these premises, we formulate a set of quantitative hypotheses that generate explicit, empirically testable predictions matched to PC9 data, which are presented and discussed below.

**Prediction 1: rare tolerant neighborhood fraction**. We define a tolerance neighborhood $\mathcal{R}_{c_0} \subset S_{c_0}$ in the drug-naïve context $c_0$ by requiring that the chromatin remodeling and oxidative stress response seminorms satisfy

$$p_{c_0,k}(x - \bar{x}_{\text{tol}}) < \sigma_k \text{ for all } k \in F_{\text{tol}},$$

where $\bar{x}_{\text{tol}}$ is the mean module score vector of experimentally identified tolerant cells, $F_{\text{tol}}$ denotes the set of tolerance-associated modules and $\sigma_k$ is one standard deviation of the corresponding module score distribution in the untreated population (Figure A). Within our framework, the fraction

$$\pi_{c_0} = \Pr(x \in \mathcal{R}_{c_0})$$

is predicted to lie in the interval

$$\pi_{c_0} \in [2 \times 10^{-3}, 5 \times 10^{-3}],$$

which matches the reported baseline DTP fraction of approximately $0.3\% \pm 0.1\%$ in PC9 populations (Sharma et al. 2010; Table 2). This prediction is not merely descriptive: if the empirically measured fraction of cells satisfying these neighborhood criteria consistently falls outside this interval across replicates, then the calibration of seminorm-based neighborhoods fails and the proposed local topology is falsified.

**Prediction 2: context-dependent reclassification of proximity**. For pairs $(x, x')$ drawn from the drug-naïve population, we define the neighborhood discrepancy

$$D_c(x, x') = \max_k p_{c,k}(x - x').$$

Using the same threshold $\varepsilon = \sigma_k$ as above, we classify pairs as "near" or "far" in a given context. Our framework predicts that when the context shifts from $c_0$ (no drug) to $c_1$ (EGFR-TKI), at least 25% of pairs that satisfy

$$D_{c_0}(x, x') < \varepsilon$$

will satisfy

$$D_{c_1}(x, x') > \varepsilon$$

(Figure B; Table 2). This quantitative reclassification reflects a rotation of relevant molecular features under drug exposure, as documented by drug-induced transcriptional reprogramming in PC9 cells (Shaffer et al. 2017). A substantially smaller reclassification rate would indicate that a fixed, context-independent notion of proximity suffices, contradicting the central motivation for context-indexed seminorms.

**Prediction 3: quantified degeneracy under survival-only observation.** We define phenotype equivalence by the criterion

$$| O_c(x) - O_c(x') | \leq 0.05,$$

and introduce the empirical degeneracy index

$$\deg_c(x) = \#\{x': | O_c(x) - O_c(x') | \leq 0.05\}.$$

Our framework predicts that the average degeneracy

$$\mathbb{E}[\deg_c]$$

increases by a factor of approximately 2 to 3 when observation is reduced from multi-marker profiling (for example survival plus pathway-specific reporters) to survival-only readouts (Figure C; Table 2). This increase quantifies the many-to-one mapping from transcriptional states to survival phenotype that has been reported for PC9 DTPs (Hangauer et al. 2017). Failure to observe a systematic increase in degeneracy under coarser observation would indicate that the observation map does not induce the expected quotient structure.

**Prediction 4: weak stabilization precedes centroid convergence.** We define the stabilization time $t_0$ of a trajectory $\gamma(t)$ as the smallest time such that

$$\gamma(t) \in N_{\text{viable}} \text{ for all } t \geq t_0,$$

where $N_{\text{viable}}$ is a neighborhood defined by viability-compatible seminorm bounds. Using time-course PC9 data, our framework predicts stabilization times in the range of 24–48 hours, reflecting early functional adaptation to drug (Figure D; Table 2). By contrast, convergence to any single transcriptomic centroid, measured for example by Euclidean distance in expression space, is predicted to occur later and with higher variance, typically at or beyond 72 hours (Shaffer et al. 2017). If centroid convergence were found to precede neighborhood stabilization systematically, the weak, neighborhood-based notion of convergence would be empirically undermined.

**Prediction 5: separating biological degeneracy from compression artifacts.** Low-dimensional embeddings of PC9 scRNA-seq data, such as PCA projections to 10–20 components, induce a baseline collapse rate of approximately 5%–10% of state pairs that appear close in embedding space despite being distant in the full space. Our framework predicts an additional, context-dependent excess collapse of comparable magnitude under EGFR inhibition (Figure E; Table 2), attributable to biological degeneracy rather than to dimensionality reduction alone. This excess collapse should disappear under gene-label randomization or context shuffling, providing a concrete null model. If no excess beyond the compression baseline is observed, then apparent degeneracy would be fully explainable by embedding artifacts rather than by the biological structure encoded in our space.

Overall, the PC9 example illustrates in concrete terms how our mathematical space can be used to analyze a real and well-quantified biological system characterized by experimentally measured single-cell states. The construction itself is not exclusively tied to EGFR-mutant NSCLC. What is specific to the PC9 system is the choice of seminorms, the definition of admissible states and the form of the observation map, all of which were calibrated using existing biological data. The underlying mathematical space remains the same. When moving to other cancer systems, biological specificity enters through different choices of local structure rather than through a different global framework. For example, in BRAF-mutant melanoma relevant seminorms could emphasize MITF–AXL transcriptional programs and natural stratification into transcriptional regimes, while admissible sets and observation maps could be defined by minimal residual disease phenotypes (Rambow et al. 2018). Still, in colorectal cancer under chemotherapy, the choice of seminorms can foreground diapause-like programs and stress responses, whereas admissibility conditions may become strongly history dependent in response to treatment-induced quiescence (Rehman et al. 2021).

Therefore, the same mathematical space can accommodate diverse cancer systems without forcing them into a single metric or representation. The space itself is fixed, while biological meaning can be introduced through context-dependent local structure. This separation allows comparisons across systems, clarifies which assumptions are shared and which are system specific, providing a common language for describing heterogeneity, degeneracy and stabilization across different cancers and treatments.

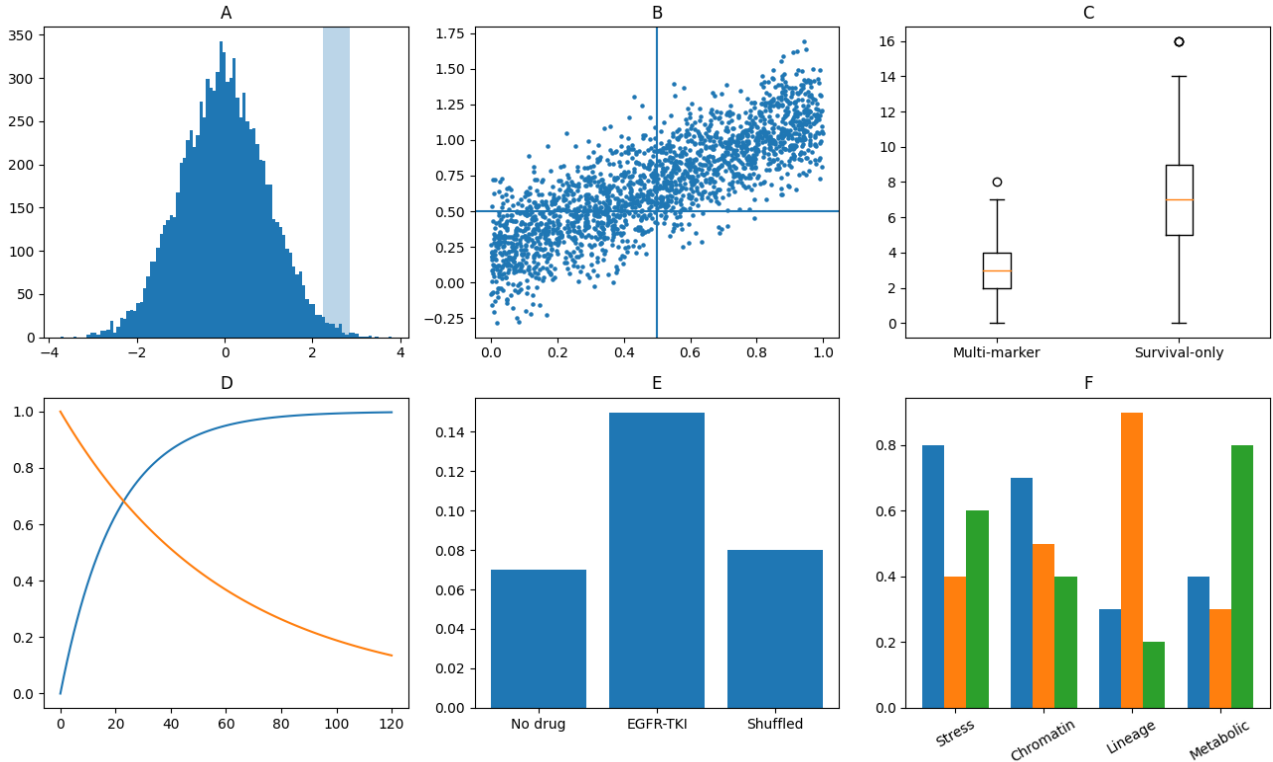

**Figure.** Schematic quantitative instantiations of predictions derived from our proposed mathematical space.
(A) Distribution of tolerance-associated module scores in the drug-naïve PC9 population. The shaded region represents the seminorm-defined tolerant neighborhood $\mathcal{R}_{c_0}$, whose fractional mass estimates the baseline frequency of drug-tolerant persister states.
(B) Context-dependent reclassification of proximity. Each point corresponds to a pair of cells, with neighborhood discrepancy in the naïve context ($D_{c_0}$) plotted against discrepancy under EGFR inhibition ($D_{c_1}$). Dashed lines indicate the common threshold $\varepsilon$; points in the lower-right quadrant identify pairs that are proximal without drug but separated under treatment.
(C) Observation-induced degeneracy. Boxplots show the distribution of the degeneracy index under multi-marker observation versus survival-only readout, illustrating the increase in many-to-one mappings from molecular state to phenotype.
(D) Weak stabilization versus pointwise convergence. Time courses compare neighborhood inclusion with centroid-based convergence, showing earlier functional stabilization than transcriptomic convergence.
(E) Separation of biological degeneracy from compression artifacts. Bar plots show state-collapse rates in low-dimensional embeddings for untreated, EGFR-TKI and shuffled controls.
(F) Generalization across cancer systems. Bar colors denote different biological systems: blue corresponds to EGFR-mutant NSCLC (PC9) orange to BRAF-mutant melanoma and green to colorectal cancer under chemotherapy. Bar heights represent the relative emphasis of selected seminorm families (stress response, chromatin remodeling, lineage programs, metabolic programs) within the same underlying mathematical space, illustrating how biological specificity is encoded by context-dependent local structure rather than by changes to the global framework.
**Note**. All panels are schematic quantitative instantiations of the predictions and do not represent empirical measurements; experimental validation requires substituting illustrative data with real datasets.

| Prediction | Quantity defined in the space | Operational measurement (PC9) | Quantitative expectation | Falsification criterion |
|---|---|---|---|---|
| P1. Rare tolerant neighborhood fraction | (Eq. P1) | Fraction of untreated cells whose chromatin and stress module scores lie within 1 SD of tolerant signature | $\pi_{c_0} \in [2 \times 10^{-3}, 5 \times 10^{-3}]$ ($\approx 0.3\% \pm 0.1\%$) | Empirical fraction consistently outside this interval across replicates |
| P2. Context-dependent reclassification of proximity | (Eq. P2) | Pairwise module-score differences before vs after EGFR-TKI | $\geq 25\%$ of pairs with $D_{c_0} < \varepsilon$ satisfy $D_{c_1} > \varepsilon$ | Reclassification rate $\ll 25\%$, indicating context-independent proximity |
| P3. Degeneracy under survival-only observation | (Eq. P3) | Size of survival-equivalence classes under different readouts | Mean degeneracy increases by factor 2–3 under survival-only observation | No systematic increase in degeneracy under coarser observation |
| P4. Weak stabilization vs centroid convergence | (Eq. P4) | Time-course scRNA-seq or lineage data | Neighborhood stabilization at 24–48 h; centroid convergence $\geq 72$ h | Pointwise centroid convergence precedes neighborhood stabilization |
| P5. Excess collapse beyond compression baseline | (Eq. P5) | PCA or latent embedding of scRNA-seq | Baseline collapse 5–10%; additional excess under drug of comparable size | Collapse fully explained by compression; no context-dependent excess |

**Table 2.** Quantitative predictions for EGFR-mutant NSCLC (PC9 system) derived from the proposed mathematical space. The mathematical expressions defining the predicted quantities are reported explicitly below.

**(Eq. P1)**
$$\pi_{c_0} = \Pr(x \in \mathcal{R}_{c_0})$$

**(Eq. P2)**
$$D_c(x, x') = \max_k p_{c,k}(x - x')$$

**(Eq. P3)**
$$\deg_c(x) = \#\{x': |O_c(x) - O_c(x')| \leq 0.05\}$$

**(Eq. P4)**
$$t_0 = \inf\{t: \gamma(t) \in N_{\text{viable}} \forall t \geq t0\}$$

**(Eq. P5)**
$$\kappa = \Pr(\|R(x) - R(x')\| < \varepsilon \land Dc(x, x') > \varepsilon')$$

## 7. CONCLUSIONS

We advance a unified mathematical approach for representing biological organization as a context-dependent space in which states, observations, admissibility and dynamics are defined coherently rather than as loosely connected modeling choices. The central idea is to treat biological states as elements of locally convex spaces indexed by context, so that proximity, stability and equivalence are defined through families of seminorms rather than through a single global metric. Context, understood as the joint effect of environment and history, is not an external parameter acting on an otherwise fixed space, but reshapes the local structure of the space itself. Within this setting, admissible sets encode biological constraints, observation maps formalize partial observability and degeneracy and directed dynamics capture irreversibility without requiring convergence to fixed points. A quantitative example illustrates how experimental measurements can be translated into these objects and how predictions emerge directly from their structure.

Elements of our mathematical structure appeared independently across disciplines. Locally convex topological vector spaces and weak topologies are classical tools in functional analysis and axiomatic field theory, where they support generalized notions of convergence and stability without reliance on global metrics (Schaefer and Wolff 1999). Fiber bundles and parameter-dependent spaces are well established in physics for encoding contextual dependence, particularly in classical mechanics and gauge theories, although context there typically parametrizes dynamics rather than reshaping the topology of space itself. Sheaf-theoretic methods have recently been introduced to formalize local data consistency, but they focus on data integration rather than on defining biological spaces with admissibility and dynamics (Hansen and Ghrist 2020). Many current models rely on metric spaces, low-dimensional embeddings or network representations, which implicitly fix what it means for two biological configurations to be close or similar. Our approach instead allows proximity to be defined locally and contextually, making explicit which biological features matter and how that relevance changes across conditions. This is useful in systems where multiple molecular configurations support the same function or where history reshapes future responses. Compared with manifold-based or embedding-based methods, our approach does not require global smoothness or dimensional reduction and therefore avoids introducing artificial notions of convergence. Compared with network-based approaches, it does not collapse space into connectivity alone, but preserves graded notions of neighborhood and admissibility. Still, stratified and hybrid spaces in control theory address regime switching and irreversibility, but usually assume fixed metric or piecewise-smooth structures. Our approach differs by integrating local convexity, admissibility, observation and directed dynamics into a single context-indexed construction tailored to biological organization.

Our approach has limitations. The construction is intentionally general, which means that concrete instantiations require careful choice of seminorms, admissible sets and observation maps, informed by domain knowledge. Poorly chosen local structures can lead to uninformative neighborhoods or trivial equivalence relations. While local convexity and weak topologies are mathematically well understood, their interpretation may be unfamiliar to biological audiences, raising a barrier to adoption. Computationally, estimating neighborhood-based quantities in high-dimensional spaces can be demanding, especially when data are sparse or noisy and empirical calibration of thresholds introduces uncertainty. Another limitation is that our approach does not specify mechanistic causal models; it provides a structural description within which causal hypotheses can be tested, but it does not replace dynamical modeling where detailed mechanisms are known. Finally, we assume that contexts can be meaningfully defined and discretized, which may be challenging in systems with continuously varying or poorly characterized environments.

Despite these limitations, our approach opens avenues for application and future research. It is applicable to systems characterized by heterogeneity, degeneracy and history dependence like cancer drug response, immune adaptation and developmental processes. Future work can refine methods for learning seminorms and admissible sets directly from data, integrate probabilistic observation maps and connect neighborhood-based stabilization with experimentally measurable timescales. Our framework generates testable experimental hypotheses, like context-dependent reclassification of proximity, quantifiable degeneracy under coarse observation and early stabilization detectable through neighborhood inclusion rather than metric convergence. These hypotheses can be evaluated using existing single-cell and longitudinal datasets, providing a direct bridge between theory and experiment. More broadly, our approach suggests recommendations for experimental design, including the importance of measuring multiple observables to control degeneracy and of explicitly accounting for history when comparing states across conditions.

In summary, we set out to define a shared mathematical space for biological processes that respects context dependence, partial observability and irreversibility without imposing inappropriate global structure. We introduced a context-fibred, locally convex space in which proximity, admissibility, observation and dynamics are defined explicitly. Many challenges in interpreting biological variability stem not from limitations of data, but from mismatches between biological phenomena and the mathematical spaces used to represent them. By reshaping the state space itself to reflect biological organization, we provide a way to formulate and test claims about stability, robustness and equivalence across diverse biological systems.

**DECLARATIONS**

**Ethics approval and consent to participate.** This research does not contain any studies with human participants or animals performed by the Author.
**Consent for publication.** The Author transfers all copyright ownership, in the event the work is published. The undersigned author warrants that the article is original, does not infringe on any copyright or other proprietary right of any third part, is not under consideration by another journal and has not been previously published.
**Availability of data and materials.** All data and materials generated or analyzed during this study are included in the manuscript. The Author had full access to all the data in the study and took responsibility for the integrity of the data and the accuracy of the data analysis.
**Competing interests.** The Author does not have any known or potential conflict of interest including any financial, personal or other relationships with other people or organizations within three years of beginning the submitted work that could inappropriately influence or be perceived to influence their work.
**Funding.** This research did not receive any specific grant from funding agencies in the public, commercial or not-for-profit sectors.
**Acknowledgements:** none.

**Authors' contributions.** The Author performed: study concept and design, acquisition of data, analysis and interpretation of data, drafting of the manuscript, critical revision of the manuscript for important intellectual content, statistical analysis, obtained funding, administrative, technical and material support, study supervision.
**Declaration of generative AI and AI-assisted technologies in the writing process.** During the preparation of this work, the author used ChatGPT 4o to assist with data analysis and manuscript drafting and to improve spelling, grammar and general editing. After using this tool, the author reviewed and edited the content as needed, taking full responsibility for the content of the publication.

# REFERENCES

1) Ahmed, K. T., S. Park, Q. Jiang, Y. Yeu, T. Hwang, and W. Zhang. 2020. "Network-Based Drug Sensitivity Prediction." BMC Medical Genomics 13 (Suppl 11): 193. https://doi.org/10.1186/s12920-020-00829-3
2) Antoine, J.-P. 2021. "Quantum Mechanics and Its Evolving Formulations." Entropy 23 (1): 124. https://doi.org/10.3390/e23010124
3) Bassett, Danielle S. and Olaf Sporns. 2017. "Network Neuroscience." *Nature Neuroscience* 20 (3): 353–64.
4) Blaszka, D., E. Sanders, J. A. Riffell, and E. Shlizerman. 2017. "Classification of Fixed Point Network Dynamics from Multiple Node Timeseries Data." Frontiers in Neuroinformatics 11: 58. https://doi.org/10.3389/fninf.2017.00058
5) Boutselis, G. I., E. N. Evans, M. A. Pereira, and E. A. Theodorou. 2021. "Leveraging Stochasticity for Open Loop and Model Predictive Control of Spatio-Temporal Systems." Entropy 23 (8): 941. https://doi.org/10.3390/e23080941
6) Breakspear, Michael. 2017. "Dynamic Models of Large-Scale Brain Activity." *Nature Neuroscience* 20 (3): 340–52.
7) Bullmore, Edward and Olaf Sporns. 2009. "Complex Brain Networks: Graph Theoretical Analysis of Structural and Functional Systems." *Nature Reviews Neuroscience* 10 (3): 186–98.
8) Burtscher, A., and L. García-Heveling. 2025. "Time Functions on Lorentzian Length Spaces." Annales Henri Poincaré 26 (5): 1533–1572. https://doi.org/10.1007/s00023-024-01461-y
9) Cassinelli, G., and P. Lahti. 2017. "Quantum Mechanics: Why Complex Hilbert Space?" Philosophical Transactions of the Royal Society A: Mathematical, Physical and Engineering Sciences 375 (2106): 20160393. https://doi.org/10.1098/rsta.2016.0393
10) Catford, J. A., J. R. U. Wilson, P. Pyšek, P. E. Hulme, and R. P. Duncan. 2022. "Addressing Context Dependence in Ecology." Trends in Ecology & Evolution 37 (2): 158–170. https://doi.org/10.1016/j.tree.2021.09.007
11) Chasapis, C. T. 2019. "Building Bridges Between Structural and Network-Based Systems Biology." Molecular Biotechnology 61 (3): 221–229. https://doi.org/10.1007/s12033-018-0146-8
12) Chen, B., and P. Miller. 2020. "Attractor-State Itinerancy in Neural Circuits with Synaptic Depression." Journal of Mathematical Neuroscience 10 (1): 15. https://doi.org/10.1186/s13408-020-00093-w
13) Cho, Y. R., and X. Hu. 2022. "Network-Based Approaches in Bioinformatics and Biomedicine." Methods 198: 1–2. https://doi.org/10.1016/j.ymeth.2021.12.010
14) Das, S., and J. R. Green. 2022. "Density Matrix Formulation of Dynamical Systems." Physical Review E 106 (5-1): 054135. https://doi.org/10.1103/PhysRevE.106.054135
15) DiFrisco, J. 2017. "Functional Explanation and the Problem of Functional Equivalence." Studies in History and Philosophy of Biological and Biomedical Sciences 65: 1–8. https://doi.org/10.1016/j.shpsc.2017.06.010
16) Del Vecchio, D. 2015. "Modularity, Context-Dependence, and Insulation in Engineered Biological Circuits." Trends in Biotechnology 33 (2): 111–119. https://doi.org/10.1016/j.tibtech.2014.11.009
17) Deneer, A., and C. Fleck. 2022. "Mathematical Modelling in Plant Synthetic Biology." Methods in Molecular Biology 2379: 209–251. https://doi.org/10.1007/978-1-0716-1791-5_13
18) Edelstein-Keshet, Leah. 2005. *Mathematical Models in Biology*. Philadelphia: SIAM.
19) Fan, L., D. Reynolds, M. Liu, M. Stark, S. Kjelleberg, N. S. Webster, and T. Thomas. 2012. "Functional Equivalence and Evolutionary Convergence in Complex Communities of Microbial Sponge Symbionts." Proceedings of the National Academy of Sciences of the United States of America 109 (27): E1878–E1887. https://doi.org/10.1073/pnas.1203287109
20) Farantos, S. C. 2024. "Hamiltonian Computational Chemistry: Geometrical Structures in Chemical Dynamics and Kinetics." Entropy 26 (5): 399. https://doi.org/10.3390/e26050399
21) Farrell, J. A., Y. Wang, S. J. Riesenfeld, K. Shekhar, A. Regev, and A. F. Schier. 2018. "Single-Cell Reconstruction of Developmental Trajectories during Zebrafish Embryogenesis." Science 360 (6392): eaar3131. https://doi.org/10.1126/science.aar3131
22) Freiberger, M. I., P. G. Wolynes, D. U. Ferreiro, and M. Fuxreiter. 2021. "Frustration in Fuzzy Protein Complexes Leads to Interaction Versatility." Journal of Physical Chemistry B 125 (10): 2513–2520. https://doi.org/10.1021/acs.jpcb.0c11068

23) Friston, Karl. 2019. "A Free Energy Principle for a Particular Physics." *Entropy* 21 (10): 940.
24) Gianni, S., M. I. Freiberger, P. Jemth, D. U. Ferreiro, P. G. Wolynes, and M. Fuxreiter. 2021. "Fuzziness and Frustration in the Energy Landscape of Protein Folding, Function, and Assembly." Accounts of Chemical Research 54 (5): 1251–1259. https://doi.org/10.1021/acs.accounts.0c00813
25) Hangauer, Matthew J., Viswanathan S. Viswanathan, Mark J. Ryan, Dominik B. Bole, James K. Eaton, Alessia Matov, et al. 2017. "Drug-Tolerant Persister Cancer Cells Are Vulnerable to GPX4 Inhibition." *Nature* 551 (7679): 247–50. https://doi.org/10.1038/nature24297.
26) Hansen, John and Robert Ghrist. 2020. "Toward a Spectral Theory of Cellular Sheaves." *Journal of Applied and Computational Topology* 4 (4): 315–58.
27) Kim, K., J. O. Yang, J. Y. Sung, J. Y. Lee, J. S. Park, H. S. Lee, B. H. Lee, Y. Ren, D. W. Lee, and S. E. Lee. 2019. "Minimization of Energy Transduction Confers Resistance to Phosphine in the Rice Weevil, Sitophilus oryzae." Scientific Reports 9 (1): 14605. https://doi.org/10.1038/s41598-019-51006-7
28) Klinger, J., and G. M. Rotskoff. 2025. "Universal Energy–Speed–Accuracy Trade-Offs in Driven Nonequilibrium Systems." Physical Review E 111 (1-1): 014114. https://doi.org/10.1103/PhysRevE.111.014114
29) Laland, Kevin N., Tobias Uller, Marcus W. Feldman, et al. 2015. "The Extended Evolutionary Synthesis: Its Structure, Assumptions and Predictions." *Proceedings of the Royal Society B* 282 (1813): 20151019.
30) Levin, Michael. 2020. "The Computational Boundary of a 'Self': Developmental Bioelectricity Drives Multicellularity and Scale-Free Cognition." *Frontiers in Psychology* 11: 149.
31) Levin, Michael. 2021. "Life, Death and Self: Fundamental Questions of Primitive Cognition Viewed through the Lens of Body Plasticity and Synthetic Organisms." *Biochemical and Biophysical Research Communications* 564: 114–33.
32) Liu, J., L. Wei, Q. Miao, S. Zhan, P. Chen, W. Liu, L. Cao, et al. 2024. "MDM2 Drives Resistance to Osimertinib by Contextually Disrupting FBW7-Mediated Destruction of MCL-1 Protein in EGFR Mutant NSCLC." Journal of Experimental & Clinical Cancer Research 43 (1): 302. https://doi.org/10.1186/s13046-024-03220-7
33) Lotfollahi, M., A. Klimovskaia Susmelj, C. De Donno, L. Hetzel, Y. Ji, I. L. Ibarra, S. R. et al. 2023. "Predicting Cellular Responses to Complex Perturbations in High-Throughput Screens." Molecular Systems Biology 19 (6): e11517. https://doi.org/10.15252/msb.202211517
34) Luebert, F., and R. A. Scherson. 2024. "Choice of Molecular Marker Influences Spatial Patterns of Phylogenetic Diversity." Biology Letters 20 (3): 20230581. https://doi.org/10.1098/rsbl.2023.0581
35) Marinazzo, D., M. Pellicoro, and S. Stramaglia. 2008. "Kernel Method for Nonlinear Granger Causality." Physical Review Letters 100 (14): 144103. https://doi.org/10.1103/PhysRevLett.100.144103
36) Minarsky, A., N. Morozova, R. Penner, and C. Soulé. 2018. "Theory of Morphogenesis." Journal of Computational Biology 25 (4): 444–450. https://doi.org/10.1089/cmb.2017.0150
37) Morabito, S., F. Reese, N. Rahimzadeh, E. Miyoshi, and V. Swarup. 2023. "hdWGCNA Identifies Co-Expression Networks in High-Dimensional Transcriptomics Data." Cell Reports Methods 3 (6): 100498. https://doi.org/10.1016/j.crmeth.2023.100498
38) Offen, C., and S. Ober-Blöbaum. 2022. "Symplectic Integration of Learned Hamiltonian Systems." Chaos 32 (1): 013122. https://doi.org/10.1063/5.0065913
39) Peidli, S., T. D. Green, C. Shen, T. Gross, J. Min, S. Garda, B. Yuan, L. J. Schumacher, J. P. Taylor-King, D. S. Marks, A. Luna, N. Blüthgen, and C. Sander. 2024. "scPerturb: Harmonized Single-Cell Perturbation Data." Nature Methods 21 (3): 531–540. https://doi.org/10.1038/s41592-023-02144-y
40) Poelwijk, F. J. 2019. "Context-Dependent Mutation Effects in Proteins." Methods in Molecular Biology 1851: 123–134. https://doi.org/10.1007/978-1-4939-8736-8_7
41) Qiu, X., Q. Mao, Y. Tang, L. Wang, R. Chawla, H. A. Pliner, and C. Trapnell. 2017. "Reversed Graph Embedding Resolves Complex Single-Cell Trajectories." Nature Methods 14 (10): 979–982. https://doi.org/10.1038/nmeth.4402
42) Rabinovich, Mikhail I., Pablo Varona, Anatol I. Selverston and Henry D. I. Abarbanel. 2008. "Dynamical Principles in Neuroscience." *Reviews of Modern Physics* 78 (4): 1213–65.
43) Rambow, Florian, Beatriz Rogiers, Othmane Marin-Bejar, Nikolas A. Aibar, Chloé Femel, et al. 2018. "Toward Minimal Residual Disease-Directed Therapy in Melanoma." *Cell* 174 (4): 843–55.e19. https://doi.org/10.1016/j.cell.2018.06.025.
44) Rehman, Sarah K., Chris Haynes, Bethany L. Collignon, et al. 2021. "Colorectal Cancer Cells Enter a Diapause-Like State to Survive Chemotherapy." *Cell* 184 (1): 226–42.e21. https://doi.org/10.1016/j.cell.2020.11.033.
45) Rood, J. E., A. Hupalowska, and A. Regev. 2024. "Toward a Foundation Model of Causal Cell and Tissue Biology with a Perturbation Cell and Tissue Atlas." Cell 187 (17): 4520–4545. https://doi.org/10.1016/j.cell.2024.07.035
46) .
47) Saxena, A., K. S. Prabhudesai, A. Damle, S. Ramakrishnan, P. Durairaj, S. Kalankariyan, A. B. et al. 2024. "A Systems Biology-Based Mathematical Model Demonstrates the Potential Anti-Stress Effectiveness of a Multi-Nutrient Botanical Formulation." Scientific Reports 14 (1): 9582. https://doi.org/10.1038/s41598-024-60112-8

48) Saxena, S., and A. Z. Rizvi. 2025. "A Novel Kernel-Based Hilbert Space Framework for Predictive Modeling of lncRNA–miRNA–Disease Interaction Networks." IEEE Transactions on Computational Biology and Bioinformatics 22 (6): 2661–2672. https://doi.org/10.1109/TCBBIO.2025.3598013
49) Schaefer, Helmut H. and Manfred P. Wolff. 1999. *Topological Vector Spaces*. 2nd ed. New York: Springer.
50) Shaffer, Steven M., Marina C. Dunagin, Sebastian R. Torborg, et al. 2017. "Rare Cell Variability and Drug-Induced Reprogramming as a Mode of Cancer Drug Resistance." *Nature* 546 (7658): 431–35. https://doi.org/10.1038/nature22794.
51) Sharma, Shankar V., Diana Y. Lee, Bin Li, et al. 2010. "A Chromatin-Mediated Reversible Drug-Tolerant State in Cancer Cell Subpopulations." *Cell* 141 (1): 69–80. https://doi.org/10.1016/j.cell.2010.02.027.
52) Song, T., G. Wang, M. Ding, A. Rodriguez-Paton, X. Wang, and S. Wang. 2022. "Network-Based Approaches for Drug Repositioning." Molecular Informatics 41 (5): e2100200. https://doi.org/10.1002/minf.202100200
53) Steinbauer, R. 2023. "The Singularity Theorems of General Relativity and Their Low Regularity Extensions." Jahresbericht der Deutschen Mathematiker-Vereinigung 125 (2): 73–119. https://doi.org/10.1365/s13291-022-00263-7
54) Stone, A., A. Youssef, S. Rijal, R. Zhang, and X. J. Tian. 2024. "Context-Dependent Redesign of Robust Synthetic Gene Circuits." Trends in Biotechnology 42 (7): 895–909. https://doi.org/10.1016/j.tibtech.2024.01.003
55) Strogatz, Steven H. 2001. "Exploring Complex Networks." *Nature* 410 (6825): 268–76.
56) Svozil, K. 2024. "(Re)Construction of Quantum Space-Time: Transcribing Hilbert into Configuration Space." Entropy 26 (3): 267. https://doi.org/10.3390/e26030267.
57) Tchitchek, N. 2018. "Navigating in the Vast and Deep Oceans of High-Dimensional Biological Data." Methods 132: 1–2. https://doi.org/10.1016/j.ymeth.2017.11.009
58) Tozzi, A. 2023. "Non-Ultrametric Phylogenetic Trees Shed New Light on Neanderthal Introgression." Organisms Diversity & Evolution: 1–9. https://doi.org/10.1007/s13127-023-00613-y
59) Tozzi, Arturo. 2025. "Dvoretzky's Theorem as a Geometric Framework for Protein Frustration." Advanced Theory and Simulations. https://doi.org/10.1002/adts.202501780
60) Verkhivker, G. M., S. Agajanian, R. Kassab, and K. Krishnan. 2022. "Landscape-Based Protein Stability Analysis and Network Modeling of Multiple Conformational States of the SARS-CoV-2 Spike D614G Mutant: Conformational Plasticity and Frustration-Induced Allostery as Energetic Drivers of Highly Transmissible Spike Variants." Journal of Chemical Information and Modeling 62 (8): 1956–1978. https://doi.org/10.1021/acs.jcim.2c00124
61) Vittadello, S. T., and M. P. H. Stumpf. 2022. "Open Problems in Mathematical Biology." Mathematical Biosciences 354: 108926. https://doi.org/10.1016/j.mbs.2022.108926
62) Wang, A., X. Li, H. Wu, F. Zou, X. E. Yan, C. Chen, C. Hu, et al. 2017. "Discovery of (R)-1-(3-(4-Amino-3-(3-chloro-4-(pyridin-2-ylmethoxy)phenyl)-1H-pyrazolo[3,4-d]pyrimidin-1-yl)piperidin-1-yl)prop-2-en-1-one (CHMFL-EGFR-202) as a Novel Irreversible EGFR Mutant Kinase Inhibitor with a Distinct Binding Mode." Journal of Medicinal Chemistry 60 (7): 2944–2962. https://doi.org/10.1021/acs.jmedchem.6b01907
63) Wang, W., D. Poe, Y. Yang, T. Hyatt, and J. Xing. 2022. "Epithelial-to-Mesenchymal Transition Proceeds through Directional Destabilization of Multidimensional Attractors." eLife 11: e74866. https://doi.org/10.7554/eLife.74866
64) Yurchenko, Sergey B. 2025. "On a Physical Theory of Causation in Multiscale Analysis of Biological Systems." *Acta Biotheoretica* 74 (1): 2. https://doi.org/10.1007/s10441-025-09511-6
65) Zhang, Y., Y. Zhang, W. Niu, X. Ge, F. Huang, J. Pang, X. Li, Y. Wang, W. Gao, F. Fan, S. Li, and H. Liu. 2021. "Experimental Study of Almonertinib Crossing the Blood–Brain Barrier in EGFR-Mutant NSCLC Brain Metastasis and Spinal Cord Metastasis Models." Frontiers in Pharmacology 12: 750031. https://doi.org/10.3389/fphar.2021.750031
66) Zhao, C., C. Su, X. Li, and C. Zhou. 2020. "Association of CD8 T Cell Apoptosis and EGFR Mutation in Non-Small Lung Cancer Patients." Thoracic Cancer 11 (8): 2130–2136. https://doi.org/10.1111/1759-7714.13504
67) Zhou, G., S. Li, and J. Xia. 2020. "Network-Based Approaches for Multi-Omics Integration." Methods in Molecular Biology 2104: 469–487. https://doi.org/10.1007/978-1-0716-0239-3_23
68) Zhu, J., C. Li, X. Peng, and X. Zhang. 2021. "RNA Architecture Influences Plant Biology." Journal of Experimental Botany 72 (11): 4144–4160. https://doi.org/10.1093/jxb/erab030